\documentclass[10pt,letterpaper,compsoc,conference]{iiswc26}

\usepackage{cite}
\usepackage{amsmath,amssymb,amsfonts}
\IfFileExists{algorithmic.sty}{\usepackage{algorithmic}}{}
\usepackage{graphicx}
\graphicspath{{figures/}}
\usepackage[dvipsnames]{xcolor}
\usepackage[final]{microtype}
\usepackage[italic]{mathastext}
\IfFileExists{libertine.sty}{\usepackage{libertine}}{}
\usepackage[T1]{fontenc}
\usepackage{textcomp}
\IfFileExists{zi4.sty}{\usepackage[varqu,varl]{zi4}}{}
\usepackage[all]{nowidow}
\usepackage[keeplastbox]{flushend}
\usepackage{fancyhdr}

\usepackage{booktabs}
\usepackage{multirow}
\usepackage{array}
\usepackage{tabularx} % column-width tables with wrapping columns
\usepackage{url}
\usepackage{hyperref}
\usepackage{cleveref}
\crefname{table}{Tab.}{Tabs.}
\Crefname{table}{Tab.}{Tabs.}
\crefname{figure}{Fig.}{Figs.}
\Crefname{figure}{Fig.}{Figs.}

\usepackage{pifont}                  % black circled numerals (\cnum)
\newcommand{\cnum}[1]{\ding{\the\numexpr#1+181\relax}}  % \cnum{1}=, ..., \cnum{8}=
\newcommand{\wcnum}[1]{\ding{\the\numexpr#1+171\relax}}  % \wcnum{1}=, ..., \wcnum{8}=

\usepackage[locale=US]{siunitx}
\usepackage{placeins}
\usepackage{subcaption}               % side-by-side subfigures (fig:stall-wave)

\usepackage[most]{tcolorbox}
\definecolor{takeawaybg}{HTML}{F2F4F7}
\definecolor{takeawayrule}{HTML}{4B5563}
\newtcolorbox{takeaway}{
  enhanced,
  colback=takeawaybg,
  colframe=takeawaybg,
  boxrule=0pt,
  leftrule=2.2pt,
  colbacktitle=takeawaybg,
  borderline west={2.2pt}{0pt}{takeawayrule},
  arc=1.2pt, outer arc=1.2pt,
  left=7pt, right=5pt, top=3pt, bottom=3pt,
  fontupper=\small,
  before upper={\textbf{\textsc{Takeaway.}}\ },
  before skip=4pt, after skip=4pt,
}

\hypersetup{
    colorlinks=true,
    linkcolor=ForestGreen,
    citecolor=ForestGreen,
    urlcolor=blue,
}

\usepackage[acronym,nopostdot,nonumberlist,nogroupskip]{glossaries}
\glsdisablehyper
\newacronym{ncu}{NCU}{Nsight Compute}
\newacronym{sol}{SOL}{Speed-of-Light throughput}
\newacronym{sm}{SM}{streaming multiprocessor}
\newacronym{tc}{TC}{tensor core}
\newacronym{bf16}{BF16}{bfloat16}
\newacronym{hbm3}{HBM3}{third-generation high-bandwidth memory}
\newacronym{gemm}{GEMM}{general matrix multiplication}
\newacronym{mfu}{MFU}{Model FLOPs Utilization}
\newacronym{moe}{MoE}{mixture of experts}
\newacronym{ffn}{FFN}{feed-forward network}
\newacronym{qkv}{QKV}{query--key--value}
\newacronym{kv}{KV}{key--value}
\newacronym{cta}{CTA}{cooperative thread array}
\newacronym{smem}{SMEM}{shared memory}
\newacronym{gmma}{GMMA}{warp-group matrix multiply-accumulate}
\newacronym{cv}{CV}{coefficient of variation}
\newacronym{slo}{SLO}{service-level objective}
\newacronym{fa3}{FA3}{FlashAttention-3}
\newacronym{rope}{RoPE}{rotary positional embedding}
\newacronym{rmsnorm}{RMSNorm}{root-mean-square normalisation}
\newacronym{tp}{TP}{tensor parallelism}
\newacronym{pp}{PP}{pipeline parallelism}
\newacronym{fp8}{FP8}{8-bit floating point}

\newcommand{\nm}[1]{\texttt{#1}}                 % short typewriter (escape _ inline)
\newcommand{\smbusy}{\nm{sm\_busy\_pct}}
\newcommand{\flashattn}{\nm{flash\_attn}}
\newcommand{\fusedmoe}{\nm{fused\_moe\_kernel}}
\newcommand{\role}[1]{\textsc{#1}}
\newcommand{\smartparagraph}[1]{\noindent\textbf{#1}}

\usepackage{titlesec}
\titlespacing*{\section}{0pt}{1.0ex plus .2ex minus .2ex}{0.6ex plus .1ex}
\titlespacing*{\subsection}{0pt}{0.8ex plus .2ex minus .2ex}{0.4ex plus .1ex}
\titlespacing*{\subsubsection}{0pt}{0.6ex plus .1ex minus .1ex}{0.3ex plus .1ex}
\usepackage{enumitem}
\setlist{nosep,leftmargin=*,itemsep=1pt,topsep=1pt,partopsep=0pt,parsep=0pt}
\begin{document}

\title{Dissecting GPU Utilization for LLM Inference on Nvidia Hopper}

\author{\IEEEauthorblockN{Mohammad Siavashi, Gerald Q. Maguire Jr.,
Dejan Kosti\'c, and Marco Chiesa}
\IEEEauthorblockA{KTH Royal Institute of Technology, Stockholm, Sweden\\
\{siavashi, maguire, dmk, mchiesa\}@kth.se}}

\maketitle
\thispagestyle{empty}
\pagestyle{empty}

%==============================================================================
\begin{abstract}
\upshape
A single ``SM utilization'' percentage can make an LLM inference workload
look compute-saturated while hiding how much useful work is being done. The
problem is not that the counter is wrong, but that it collapses several
different mechanisms into one number. This is most severe during decode,
where each request contributes only one new token and dense projection GEMMs
become small-row matrix multiplications. On Hopper, the bfloat16 GMMA path
executes these operations in fixed 64-row matrix fragments, so small-batch
decode can fill only a small fraction of each fragment with real token rows.

In this paper, we profile vLLM with FlashAttention-3 and cuBLASLt on an
H100 NVL across cold prefill, warm prefill, and decode, sweeping sequence
length and batch size. We replace the usual single utilization number with
eight counter-validated views derived from raw Nsight Compute reports, each
pinned to an NCU counter or explicit formula. Together, these views map
utilization gaps to concrete mechanisms---fragment fill, occupancy limits,
stall signatures, wave quantization, and kernel selection---across four
production models and six per-layer kernel roles.

% It shows, for instance, that production
% range decode (batch sizes 1--8) fills only 1.6--12.5\% of GMMA
% M-fragment rows, averaging 5.86\% over 136 configurations, so reported
% compute throughput can overstate useful decode FLOP throughput by 8--64
% times; that FlashAttention-3 decode is barrier-stall- rather than
% memory-stall-bound; and that dense bfloat16 occupancy is capped jointly
% by registers and shared memory. Improving decode therefore requires
% raising the effective matmul row count---through cross-request token
% packing or persistent-decode kernels---and reporting multiple
% counter-pinned views rather than one SM-utilization number.
\end{abstract}

% Reset acronym first-use tracking so the body re-introduces
% each acronym in full on its own first occurrence (per IEEE
% conference convention; abstract is typically read separately).
\glsresetall


%==============================================================================
\section{Introduction}

A GPU can look busy for reasons that have little to do with useful LLM
work.  That distinction now matters in practice: LLM inference consumes
enough hyperscale GPU time that a utilization number quickly becomes a
capacity-planning, batching, and optimization decision.  Yet the
question ``is the GPU being used?'' is often answered with a single
percentage: NVIDIA \gls{ncu}'s
\texttt{sm\_\_throughput.}\allowbreak\texttt{avg.}\allowbreak\texttt{pct\_of\_peak\_sustained\_elapsed},
conventionally called \emph{SM utilization} or \smbusy{}.  This counter
is useful, but narrower than the name suggests.  It is an elapsed-cycle
pipeline-throughput proxy, not the \texttt{sm\_\_cycles\_active}
``$\geq$1 warp resident'' counter, not an occupancy measurement, and
not a direct measure of useful matrix-multiply throughput.  The problem
is not that the counter is wrong; it is that a single scalar is being
asked to explain too many different denominators.

This ambiguity is concrete.  In Qwen3-14B dense decode on an H100 at
batch size 32, the GEMMs report ${\approx}72\%$ \smbusy{}, reading as
near compute-saturation---yet only half of each Hopper \texttt{m64}
matrix fragment holds real token rows, and \texttt{long\_scoreboard}
memory stalls still dominate the issue budget; the cause is
architectural, an \texttt{m64} fragment floor in Hopper's \gls{bf16}
\gls{gmma} path that no tile retuning can repair (\S\ref{sec:tc}).
The same ambiguity appears the moment the serving regime changes.
For the same projection role at the same model, context length $S$, and
batch size $B$, \smbusy{} can be $\sim$97\% in cold prefill, where
\texttt{cuBLASLt}'s autotuner selects a large tile, yet only $\sim$9\%
in warm prefill after a cached prefix collapses the row dimension
toward decode size (\S\ref{sec:tc}).  Even occupancy can appear to tell two stories at once: the
same kernel is only $\sim$14\% occupied against the architectural
64-warp ceiling, but 75--95\% occupied against its own
register/shared-memory resource cap (\S\ref{sec:two-denoms}).  Each
reading is internally consistent.  The confusion comes from forgetting
which denominator each metric uses.

This paper makes those denominators explicit.  We instrument vLLM with
\gls{fa3}, \texttt{cuBLASLt}, and Hopper's \gls{gmma} path on an H100
NVL with 94\,GB \gls{hbm3}~\cite{h100nvl}.  We study four production
models: Meta-Llama-3-8B, Qwen3-14B, Qwen3-32B, and Qwen3-30B-A3B, the
last a sparse \gls{moe} model with top-8 routing over 128 experts.  For
each model, we sweep three serving regimes---cold prefill, warm
prefill, and decode---along with context length $S$, batch size $B$,
and per-layer kernel role: QKV projection, attention output
projection, FFN Gate/Up projection, FFN Down projection, attention,
and the MoE expert path where applicable.  For every
(model, regime, $S$, $B$, role) point, we compute eight utilization
views directly from raw Nsight Compute report files, pinning each view
either to an NCU counter or to an explicit formula.

The central result is that decode underutilization is not one
bottleneck but several---resource limits, spatial wave quantization,
and an instruction-set fragment floor---each with a different fix, and
each invisible to the single scalar.  Existing inference profiling often reports
either end-to-end latency or a single utilization axis, with no
methodology to say whether a low number comes from memory stalls, tile
shape, CTA wave loss, an occupancy cap, attention-side synchronization,
or an instruction-set fragment floor.  Our goal is therefore not
another aggregate benchmark, but a counter-pinned map from the
symptom---a high or low utilization reading---to the mechanism that
produced it.  The contributions below tie each mechanism to its
measurement.

\smartparagraph{Scope.}  Single H100 NVL, BF16, TP=1, unmodified
\texttt{cuBLASLt}/\gls{fa3}/vLLM, and synthetic prefix-cache states
(near-perfect hit on warm prefill).  We do not claim the absolute
values transfer unchanged to other H100 SKUs, tensor-parallel
configurations, FP8 paths, or production traces with intermediate
prefix-cache hit rates; those limitations are revisited in
\S\ref{sec:threats}.

\smartparagraph{Contributions.}
\wcnum{1}~\textbf{\emph{A reusable eight-metric, counter-validated
reporting methodology}} that replaces the single SM-utilization
number with eight complementary views
(\S\ref{sec:characterization}), each --- together with its derived
quantities --- pinned to an NCU counter or explicit formula with
verified semantics (Tab.~\ref{tab:metrics}).
\wcnum{2}~\textbf{\emph{A per-role, per-regime characterization}} of the
vLLM\,+\,\gls{fa3}\,+\,\texttt{cuBLASLt} serving stack,
across cold prefill, warm prefill, and decode on four production
models, with every finding tied to a mechanism: dense-GEMM occupancy,
under \texttt{cuBLASLt}'s nvjet tile selection, is co-limited by
registers \emph{and} shared memory, so a one-sided
register cut cannot deepen the warp pool (\S\ref{sec:rs-bind});
\gls{fa3} and dense decode diverge in stall signature, \gls{fa3}'s
leading stall shifting away from memory dependency toward barrier and wait
(\S\ref{sec:stalls}); decode-QKV launch grids lose chip coverage to
wave quantization (\S\ref{sec:wave}); and warm prefill collapses
toward decode via the warm-cache tile mechanism (\S\ref{sec:tc}).
\wcnum{3}~\textbf{\emph{A hardware-grounded ceiling on small-batch
decode matmul}}: Hopper's \gls{gmma} \texttt{m64} fragment caps every
decode dense-GEMM at 1.6--12.5\% fragment fill, so a per-kernel
compute-throughput reading overstates useful matmul
throughput by 8--64$\times$ --- a measurement-attribution gap, not
recoverable headroom; the lever is to raise the row
dimension $M$ (persistent-decode kernels, cross-request token
packing), not to retile (\S\ref{sec:tc}).
\wcnum{4}~\textbf{\emph{Controlled experiments and an artifact
release}}: a \texttt{cuBLASLt} tile-space enumeration showing no
tile achieves full SM coverage for decode-QKV (\S\ref{sec:wave}),
and a $B{=}32$ validation arm pinning the compute--memory crossover
(\S\ref{sec:b32}).
We plan to make the raw NCU and Nsys reports and the derived CSV datasets
publicly available.

%==============================================================================
\section{Background and Motivation}\label{sec:bg}

\subsection{The H100 microarchitectural budget}

Our NVIDIA H100 NVL % (94\,GB HBM3, PCIe Gen 5)~\cite{h100nvl}
has 132 \glspl{sm}.  Each SM contains four sub-partitions (warp schedulers),
each with one 4th-generation
\gls{tc} --- 528 \glspl{tc} in total~\cite{wgmma}.  Peak dense
\gls{bf16} \gls{tc} throughput is 835~TFLOP/s (\num{1671}~TFLOP/s
with 2:4 structural sparsity, which we do not exploit)~\cite{h100nvl}.
The \gls{bf16} \gls{gmma} instruction (\texttt{wgmma.mma\_async}) has a
fixed 64-row $M$-axis fragment~\cite{wgmma}, an instruction-set property
we analyze as a small-batch decode ceiling in \S\ref{sec:tc}.
Peak \gls{hbm3} bandwidth is 3.9~TB/s~\cite{h100nvl}.  The L2 cache
is 60~MB unified across \glspl{sm}.  The per-SM resource budget is
\num{65536} 32-bit registers, 228\,KB shared memory (opt-in max),
64 resident warps (architectural ceiling), 32 resident thread blocks,
and 4 Tensor Cores.
Under the roofline model~\cite{roofline}, the arithmetic-intensity
ridge is \(I^{*} = 835\,/\,3.9 \approx 214\) FLOP/byte in BF16.
Kernels below \(I^{*}\) are HBM-bandwidth-bound; kernels above
\(I^{*}\) are compute-bound.

\subsection{Kernel families in an LLM inference layer}

A standard transformer layer in vLLM dispatches the following dense
kernel roles per call (\Cref{tab:roles}): a \gls{qkv} projection, an
output projection (\texttt{o\_proj}), a Gate/Up projection, a Down
projection, and \gls{fa3} for attention.  The dense BF16
\gls{gemm} kernels are dispatched by \texttt{cuBLASLt}, built on
CUTLASS~\cite{cutlass}, and typically come from the
\texttt{nvjet\_hsh\_*} family on Hopper.

\gls{fa3}~\cite{flashattn3} launches as a \emph{persistent} kernel
(\texttt{FlashAttnFwdSm90}, grid \(132 \times 1 \times 1\)): it
launches one block per SM and loops over attention tiles internally,
so chip coverage remains full regardless of problem size
(\S\ref{sec:wave}).  It is followed by a smaller combine kernel
(\texttt{FlashAttnFwdCombine}, grid 5--80 \glspl{cta}).

On the sparse MoE Qwen3-30B-A3B, the \gls{ffn} role is replaced by a
fused grouped-GEMM (\fusedmoe) that processes the top-8 of 128 experts
in a single launch, preceded by a tiny ($\sim$6\,$\mu$s, 2-CTA)
router-gate projection.  Two auxiliary kernels are tracked but not analyzed individually --- \role{cache\_w}
(\texttt{reshape\_and\_cache\_flash}, $<$2\% of decode) and
\role{silu\_mul} (FFN SiLU); hence, 7 dense (8 MoE) rows per
cell in Tab.~\ref{tab:coverage}.

\begin{table}[t]
\centering
\caption{Per-layer kernel roles tracked in this study. Dense
models use the top five rows; MoE models replace Gate/Up$+$Down
with \role{moe}.  % Two auxiliary kernels are also tracked but not analyzed individually --- \role{cache\_w} (\texttt{reshape\_and\_cache\_flash}, $<$2\% of decode) and \role{silu\_mul} (FFN SiLU) --- giving the 7 dense (8 MoE) rows per cell in Tab.~\ref{tab:coverage}.
}
\label{tab:roles}
\footnotesize
\setlength{\tabcolsep}{3pt}
\begin{tabular}{@{}p{0.16\linewidth}>{\raggedright\arraybackslash}p{0.78\linewidth}@{}}
\toprule
Role & What it captures \\
\midrule
\role{qkv}         & Combined Q/K/V projection (dense \gls{gemm}). \\
\role{o\_proj}     & Attention-output projection (dense \gls{gemm}). \\
\role{gate\_up}    & FFN Gate/Up projection (dense, fused for SiLU). \\
\role{down\_proj}  & FFN Down projection (dense \gls{gemm}). \\
\role{flash\_attn} & \gls{fa3} persistent main kernel; the combine kernel is reported under \role{flash\_combine}. \\
\midrule
\role{moe}          & MoE-only. Fused grouped-GEMM (\fusedmoe) plus the small \texttt{moe\_align}\,/\texttt{moe\_sum}\,/scatter--gather kernels that surround it; carries the FFN compute for MoE models. The per-layer router-gate \gls{gemm} (hidden\,$\to$\,$E$ experts) is tracked as a separate \role{router} row (counted in \Cref{tab:coverage}; bucketed into the non-FFN ``dense'' column of \Cref{tab:b32}); the \texttt{topkGatingSoftmax} selection kernel is profiled in the CSV cache but not reported. \\
\bottomrule
\end{tabular}
\end{table}

\subsection{Three operational regimes}\label{sec:regimes}

Three regimes exercise the H100 differently.
\cnum{1}~\textbf{Cold prefill}: every prompt is processed end-to-end
with no prior \gls{kv} cache reuse; dense GEMMs see row dimension
\(M\!\approx\!S{\cdot}B\) and arithmetic intensity is high.
\cnum{2}~\textbf{Warm prefill}: only a short suffix is re-evaluated
after a cached prefix is reused, collapsing \(M\) from
\(\approx\!S{\cdot}B\) toward \(\approx\!B\); representative of
multi-turn chat.
\cnum{3}~\textbf{Decode}: one query token per request per step
against a growing KV cache (\(M\!\approx\!B\)); the dominant
production regime; HBM-bandwidth-bound.

\smartparagraph{How the regimes are realised.}
vLLM's \emph{prefix cache}~\cite{vllm,sglang} reuses already-computed
KV blocks across requests (or warmup iterations) sharing a prefix;
this flag alone controls the two prefill regimes.  A
\textbf{cold-prefill} cell runs with
\texttt{enable\_prefix\_caching=false} (0\% hit by construction),
so dense GEMMs see the full \(M\!\approx\!S{\cdot}B\) row dimension.
A \textbf{warm-prefill} cell enables the cache after an
identical-prompt warmup; with vLLM's 16-token cache blocks, the
profiled call re-evaluates at most one trailing 16-token block, a
deterministic $\geq$99.2\% hit at $S{=}2048$ rising to
$\geq$99.9\% at $S{=}16384$ (decode-sized \(M\)).  The two prefill
regimes therefore bracket the cache-reuse axis at its endpoints; the
near-perfect-hit warm case is the small-payload limit, distinct from
a realistic intermediate-hit multi-turn workload (\S\ref{sec:threats}).

Decode is profiled at \texttt{max\_tokens=2} with the leading prefill
iteration discarded \textit{post hoc}.  In every regime, at least three
warmup iterations precede the profiled one and CUDA graphs are
captured ahead of time.

\subsection{Occupancy has two denominators}\label{sec:two-denoms}

Achieved occupancy can be normalized against two denominators that
often differ by about $5\times$; this is one of the most common and
consequential confounds in interpreting GPU \textit{utilization}.
\wcnum{1}~\textbf{Architectural limit}: 64 warps/SM, a hardware constant
independent of the kernel.
\wcnum{2}~\textbf{Kernel resource cap} (NCU's ``Theoretical Active Warps per
SM''): the maximum warps the kernel's resources permit, defined as the
minimum over register file, \gls{smem}, max blocks, and named barriers;
typically 8--12 for the kernels in our sweep, far below 64.

\gls{fa3} main is the cleanest illustration: it reaches 7.6 warps/SM,
which is 12\% of the architectural limit yet 95\% of its 8-warp
resource cap, i.e.\ near saturation. Both readings are correct but
answer different questions; keeping the normalization explicit resolves
much practitioner confusion.

%==============================================================================
\section{Methodology}\label{sec:method}

\subsection{Profiling pipeline}

All measurements use NVIDIA Nsight Compute~2024.3 (\texttt{ncu})~\cite{ncu-doc}.
We invoke \texttt{ncu} per kernel with the \texttt{MemoryWorkloadAnalysis},
\texttt{LaunchStats}, \texttt{SpeedOfLight}, \texttt{WarpStateStats}, and
\texttt{Occupancy} sections enabled.
Per-layer, per-role kernel instances (32, 40, 64, and 48 layers for
Llama-3-8B, Qwen3-14B, Qwen3-32B, and Qwen3-30B-A3B respectively) are
aggregated using NCU's per-kernel CSV export.

Three metric families (per-stall-reason counters, raw tensor-pipe counters,
and the per-call \gls{fa3} main-vs.-combine split) are renamed or blended in the
upstream CSV, so we re-read the underlying \texttt{.ncu-rep} files into
three caches: \texttt{h100\_role\_stall\_cache.csv} (per-(model, role, regime,
$S$, $B$) stall counters), \texttt{h100\_role\_tc\_cache.csv} (tensor-core counters
from \texttt{sm\_\_pipe\_tensor\_*}), and \texttt{h100\_fa\_split\_cache.csv}
(\gls{fa3} main and combine separated).
NCU reports metrics per kernel launch.  For utilization metrics we average
matched calls across a model's layers; for time decomposition we sum across
roles within a token.  Both views are required: a single per-token aggregate
can hide kernel-launch overhead.

\subsection{Metrics and their denominators}

\Cref{tab:metrics} defines each metric via its NCU counter or explicit formula.
A recurring confound is that nominally similar ``utilization'' quantities use
\emph{different denominators}: SM utilization is elapsed-cycle (not per-active-SM);
the two occupancy readings normalize by the 64-warp architectural limit vs.\ the
kernel's own resource cap; and compute SOL is reported both per active SM and
device-wide ($\times\,\text{SMs reached}/132$, the \gls{mfu}-like
form~\cite{palm}), the two coinciding only when a launch reaches all 132 SMs.

\begin{table*}[t]
\centering
% tighten float + tabular spacing a bit
\setlength{\abovecaptionskip}{2pt}
\setlength{\belowcaptionskip}{0pt}
\setlength{\tabcolsep}{3pt}
\renewcommand{\arraystretch}{0.92}
\caption{The eight primary metrics (\S\ref{sec:characterization})
and their derived quantities. Some metrics occupy two rows where
distinct denominators or normalizations apply.}
\label{tab:metrics}
\begin{tabular}{>{\raggedright\arraybackslash}p{0.18\linewidth}>{\raggedright\arraybackslash}p{0.40\linewidth}>{\raggedright\arraybackslash}p{0.36\linewidth}}
\toprule
Metric & NCU counter / formula & Interpretation \\
\midrule
SM utilization & \texttt{sm\_\_throughput.}\allowbreak\texttt{avg.}\allowbreak\texttt{pct\_of\_peak\_sustained\_elapsed} (NCU \emph{Compute (SM) Throughput} section metric) & Fraction of peak-sustained SM-pipe throughput over elapsed cycles (max over SM sub-pipelines, ideal-SMSP load balance); a time-utilization proxy, \emph{not} an occupancy, \texttt{sm\_\_cycles\_active}, or ``$\geq$1 warp active'' counter. \\
Compute SOL (per active SM) & \texttt{sm\_\_throughput.}\allowbreak\texttt{avg.}\allowbreak\texttt{pct\_of\_peak\_sustained\_elapsed} (the same counter as ``SM utilization'' above) & Peak-sustained SM-pipe throughput read on the SMs the kernel reaches; a duty-cycle proxy for matmul throughput on dense-GEMM kernels. \\
Compute SOL (device-wide) & $\text{compute\_sol\_pct} \cdot \text{SMs\_reached}/132$ & The same proxy rescaled by spatial coverage and credited against all 132 SMs; a per-kernel analog of model-level \gls{mfu}.\\
\cmidrule(lr){1-3}
SMs used & $\min(N_\text{CTA}, 132)$ from launch grid & Spatial chip coverage. \\
\cmidrule(lr){1-3}
Achieved warps/SM & NCU \emph{Achieved Active Warps Per SM} $=$ \texttt{sm\_\_warps\_active.}\allowbreak\texttt{avg.}\allowbreak\texttt{per\_cycle\_active} (over active SMs; achieved occupancy $=$ \texttt{sm\_\_warps\_active.}\allowbreak\texttt{avg.}\allowbreak\texttt{pct\_of\_peak\_sustained\_active}) & Resident warp depth. \\
Occupancy vs.\ architectural limit & warps/SM $\div$ 64 & Saturation vs.\ the SM's hardware ceiling (64 warps). \\
Occupancy vs.\ kernel resource cap & warps/SM $\div$ (NCU ``Theoretical Active Warps per SM'') & Saturation vs.\ the most warps this kernel's registers/SMEM allow. \\
\cmidrule(lr){1-3}
Stall budget & \texttt{smsp\_\_average\_warps\_issue\_stalled\_}\allowbreak$\langle r\rangle$\allowbreak\texttt{\_per\_issue\_active.ratio} per reason $r$, normalized by \texttt{warp\_cycles\_per\_issued} & Why resident warps fail to issue (per-reason share of the per-issued stall budget). \\
\cmidrule(lr){1-3}
L2 fit ratio $\rho$ & footprint / 60~MB & Per-call working set vs.\ H100 L2. \\
\cmidrule(lr){1-3}
Wave efficiency & $N_\text{CTA} / (\lceil N_\text{CTA}/132 \rceil \cdot 132)$ & Launch-grid tiling efficiency. \\
\cmidrule(lr){1-3}
\gls{gmma} $M$-fragment fill & $\eta_M = M_\text{actual} / (\lceil M_\text{actual}/64 \rceil \cdot 64)$; the m64 floor is the BF16 \texttt{wgmma.mma\_async} architectural minimum~\cite{wgmma} & Per-instruction M-axis tile fill; capped above by the WGMMA hardware fragment (\S\ref{sec:tc}). \\
\cmidrule(lr){1-3}
Analytical launch overhead & $\frac{N_\text{kern}\,\tau_\text{launch}}{N_\text{kern}\,\tau_\text{launch} + \sum_k t_k}$, $\tau_\text{launch}{=}5\,\mu\text{s}$ & Share of iteration spent on CPU$\to$GPU dispatch (lower bound). \\
Kernel-selection diversity & $|\{\text{kernel\_name}(s,b)\}|$ across the sweep & Number of distinct kernels the autotuner picks; a stability surrogate. \\
\bottomrule
\end{tabular}
\end{table*}

\subsection{Working-set model}
\label{sec:working-set}

For the L2-fit analysis, we estimate the per-kernel-call footprint (BF16 bytes) as
$\text{footprint} = W_\text{weights} + A_\text{act} + K_\text{kv}$, where $W_\text{weights}$ is weight bytes read, $A_\text{act}$ is activation bytes read/written, and $K_\text{kv}$ is KV-cache bytes read/written.
Notation: $H$ hidden size; $n_\text{attn}$ attention heads; $n_\text{kv}$ KV heads; $d_\text{head}$ head dim; $H_q = n_\text{attn} d_\text{head}$; $H_{kv} = n_\text{kv} d_\text{head}$; $F$ FFN intermediate; $S$ context length; $B$ batch size; $T = S\cdot B$ (prefill) or $T = B$ (decode); \hbox{$k$ experts/token}.

%==============================================================================
\section{Workloads, Systems, and Measurement}\label{sec:workload}

\smartparagraph{Hardware and software.}
We run vLLM (BF16, TP=1) on an H100 NVL %(94\,GB HBM3, PCIe Gen 5)
with \gls{fa3} enabled, the \texttt{cuBLASLt} Hopper GMMA path, and
CUDA Graphs enabled.  Graphs stay on for the whole run, including the
end-to-end timings we report; the only exception is the NCU-profiled
iteration itself, which the profiler requires to be uncaptured so that
per-kernel sections remain valid.  No wall-clock claim in this paper is
taken from a profiled iteration.  We run three warmup iterations before
each profiled iteration.  Model
architectural dimensions are taken from
public HuggingFace \texttt{config.json} files and listed in
\Cref{tab:models}.

\begin{table}[t]
\centering
\caption{Model architectures from HuggingFace
\texttt{config.json}~\cite{llama3,qwen3}. Qwen3-30B-A3B's $F$ is per-expert.}
\label{tab:models}
\small
\setlength{\tabcolsep}{3.5pt}
\renewcommand{\arraystretch}{0.95}
\resizebox{\columnwidth}{!}{%
\begin{tabular}{lrrrrrrl}
\toprule
Model & \(H\) & \(n_\text{attn}\) & \(n_\text{kv}\) & \(d_\text{head}\) & \(F\) & \(L\) & Type \\
\midrule
Meta-Llama-3-8B & 4096 & 32 & 8 & 128 & 14336 & 32 & Dense \\
Qwen3-14B       & 5120 & 40 & 8 & 128 & 17408 & 40 & Dense \\
Qwen3-32B       & 5120 & 64 & 8 & 128 & 25600 & 64 & Dense \\
Qwen3-30B-A3B   & 2048 & 32 & 4 & 128 & 768\,(per-expert) & 48 & Sparse MoE (128 experts, top-8) \\
\bottomrule
\end{tabular}%
}
\end{table}

\smartparagraph{Configuration sweep.}
For each model we sweep \((S,B)\) corners; coverage is intentionally
asymmetric across regimes.  Cold prefill is swept densely on all four
models.  Warm prefill is profiled at one small corner and two large
memory-limit corners per model: because the warm-cache state is
shape-insensitive (\S\ref{sec:regimes}), a few corners suffice to
bracket it.  Decode is swept across \(B{\in}\{1,2,4,8\}\) to expose
compute--memory crossover trends; Qwen3-32B decode is the one
KV-memory-capped case (256\,KB/token forces no \(B{=}8\), and long
\(S\) admits only \(B{\le}4\)).

\Cref{tab:coverage} enumerates the main profiled grid.  A \emph{cell}
is one profiled $(\text{model}, \text{regime}, S, B)$ corner; a
\emph{row} is one per-role NCU line within a cell, i.e.\ a
$(\text{model}, \text{regime}, S, B, \text{role})$ tuple, after
per-layer aggregation (7 roles for dense models; 8 for
Qwen3-30B-A3B, which adds router and \fusedmoe{} and has no separate
down-proj role).  Warm prefill may additionally expose a
\role{flash\_combine} role (\gls{fa3} combine; otherwise aggregated into
\role{flash\_attn}), so warm-prefill rows are not always a strict
cells\,$\times$\,roles product.  The $B{=}32$ validation arm
(\S\ref{sec:b32}) adds four decode cells, one per model
($B{=}4$ for the KV-memory-capped Qwen3-32B).

\begin{table}[t]
\centering
\caption{Configuration-sweep coverage.}
\label{tab:coverage}
\setlength{\tabcolsep}{2pt}
\scriptsize
\resizebox{0.9\columnwidth}{!}{%
\begin{tabular}{@{}llrr@{}}
\toprule
Regime & Model / swept $(S\!\times\!B)$ corners & Cells & Rows \\
\midrule
\multicolumn{4}{@{}l}{\emph{Cold prefill} (prefix caching off)} \\
 & Llama-3-8B\; $S\{1024..8190\}\!\times\!B\{1,2,4\}{+}(8190,8)$ & 13 & 91 \\
 & Qwen3-14B\;\; $S\{1024..8192\}\!\times\!B\{1,2,4\}{+}(16384,4),(16384,8)$ & 14 & 98 \\
 & Qwen3-30B-A3B\;\; $S\{1024..8192\}\!\times\!B\{1,2,4\}{+}(8192,8)$ & 13 & 104 \\
 & Qwen3-32B\;\; 12 of $S\{1024..16384\}\!\times\!B\{1,2,4\}$    & 12 & 84 \\
\cmidrule(l){2-4}
 & \emph{subtotal}                                               & \textbf{52} & \textbf{377} \\
\midrule
\multicolumn{4}{@{}l}{\emph{Warm prefill} (prefix caching on, $\geq$99\% hit)} \\
 & Llama-3-8B $(2048,1),(8190,4),(8190,8)$                       & 3 & 22 \\
 & Qwen3-14B $(2048,1),(16384,4),(16384,8)$                      & 3 & 22 \\
 & Qwen3-30B-A3B $(2048,1),(8192,4),(8192,8)$                    & 3 & 25 \\
 & Qwen3-32B $(2048,1),(16384,2),(16384,4)$                      & 3 & 21 \\
\cmidrule(l){2-4}
 & \emph{subtotal}                                               & \textbf{12} & \textbf{90} \\
\midrule
\multicolumn{4}{@{}l}{\emph{Decode}} \\
 & Llama-3-8B $S\{2048..8190\}\!\times\!B\{1,2,4,8\}$            & 12 & 84 \\
 & Qwen3-14B $S\{2048..8192\}\!\times\!B\{1,2,4,8\}{+}(16384,8)$ & 13 & 91 \\
 & Qwen3-30B-A3B $S\{2048..8192\}\!\times\!B\{1,2,4,8\}$         & 12 & 96 \\
 & Qwen3-32B $(2048,1),(16384,2),(16384,4)$                      & 3  & 21 \\
\cmidrule(l){2-4}
 & \emph{subtotal}                                               & \textbf{40} & \textbf{292} \\
\midrule
 & \textbf{Total}                                                & \textbf{104} & \textbf{759} \\
\bottomrule
\end{tabular}%
}

\vspace{1pt}
{\scriptsize Llama-3-8B uses $S{=}8190$ (2 tokens below its 8192 limit, for decode headroom).}
\end{table}

%==============================================================================
\section{The Eight-Metric Characterization}\label{sec:characterization}

The eight metrics are described in the order we recommend for
diagnosing an inference-utilization issue.

\subsection{SM time-utilization}

\begin{figure}[t]
\centering
\includegraphics[width=\linewidth]{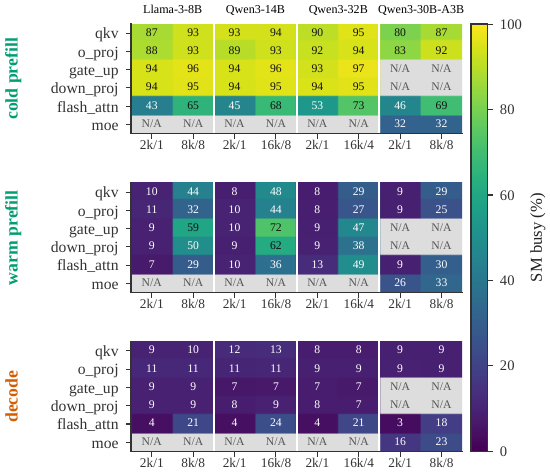}
\caption{SM utilization heatmaps by model, regime, role, and $(S,B)$. Rows are kernel roles; columns are swept $(S,B)$ corners; one panel per (model, regime).}
\label{fig:smbusy}
\end{figure}

\Cref{fig:smbusy} reports raw \smbusy{} as a heatmap; the ranges below
span all four models and all swept $(S,B)$ corners.

\smartparagraph{Cold prefill.}
Dense GEMMs saturate the issue pipeline (73--97\%, mean 92\%; peak
97.4\% on Llama-3-8B \role{down\_proj} at $S{=}8190, B{=}2$).
Qwen3-30B-A3B's FFN (reported under \role{moe}) sits at 30--32\%.
\gls{fa3} spans 27--73\%, rising with $S$.

\smartparagraph{Warm prefill.}
Utilization drops because \texttt{cuBLASLt}'s warm-cache selects smaller
tiles: dense GEMMs collapse to 8--11\% at $(S{=}2048, B{=}1)$ and
recover unevenly at the larger corner (14--73\%, mean 33\%).
Qwen3-32B's $(16384,4)$ warm cell returns to GEMM-like utilization while
the others stay decode-like; \gls{fa3} is 7--49\%; MoE is 26--33\%.

\smartparagraph{Decode.}
Decode is lowest and relatively uniform: dense 6.5--12.7\% (mean 9.4\%),
\gls{fa3} 3.4--23.9\% (growing with KV working set), and MoE 16--23\%.
Thus, ``decode under-utilizes the GPU'' is correct but coarse: small-payload
warm prefill is already decode-like, and the cold\,$\to$\,warm drop is itself
roughly $4\times$ on the dense-GEMM mean.  The next seven metrics isolate the
mechanisms.

%\smartparagraph{Model size and decode-QKV coverage.}
%Decode-QKV \smbusy{} is non-monotonic in parameter count (Llama-3-8B
%9--11\%, Qwen3-14B 12.2--12.7\%, Qwen3-32B 8.1--8.2\%):
%\texttt{cuBLASLt} tile choice depends on $(M,N,K)$, not parameter
%count.

\subsection{SM coverage and warp-pool occupancy}\label{sec:warp-pool}

\begin{figure}[t]
\centering
\includegraphics[width=0.95\linewidth]{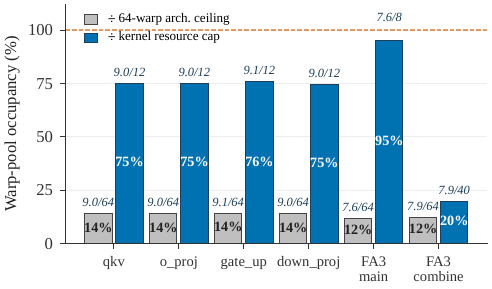}
\caption{Achieved warps/SM under two denominators on the canonical cell (Llama-3-8B decode, $S{=}2048, B{=}1$): normalized by the 64-warp architectural limit and by each kernel's resource cap $\eta_\text{cap}$ (12 dense, 8 \gls{fa3} main).  Dense decode values stay within 9.0--9.4~warps/SM across the full sweep ($S\,{\in}\,\{2\text{k}\dots16\text{k}\}$, $B\,{\in}\,\{1\dots8\}$, 8B--32B models): the empty warp pool is not a small-budget artifact, and this cell is representative.}
\label{fig:warp-pool}
\end{figure}

\Cref{fig:warp-pool} reframes utilization along two additional axes.
\smartparagraph{SM coverage.}
Across the decode sweep, dense projections launch on 80--132 of
132 SMs.  QKV grid sizes are model-specific (Llama-3-8B 96 CTAs;
Qwen3-14B 132, full; Qwen3-32B and Qwen3-30B-A3B 80; see
\S\ref{sec:wave} and \Cref{fig:wave}).  \role{o\_proj} and
\role{down\_proj} land at 128; \role{gate\_up} lands at 92--132.
Thus, limited ``SMs reached'' is not the dominant decode issue:
most SMs do receive work.

\smartparagraph{Warp-pool depth.}
Achieved warps/SM is essentially flat across the sweep ---
$\approx$8.9--9.5 for dense BF16 GEMMs (cold prefill, warm prefill,
and decode agree) and 7.6 for \gls{fa3} main in decode.  Normalized by the
64-warp architectural limit, these correspond to 12--15\% (i.e.,
``the warp pool is mostly empty'').  Normalized instead by each
kernel's resource cap (12 for the dense GEMMs, 8 for \gls{fa3} main), they
are 75--95\%, i.e., at or near the kernel's own ceiling.
This $\approx 5{\times}$ denominator gap from \S\ref{sec:two-denoms}
holds for every dense decode kernel in our sweep (\Cref{fig:warp-pool});
it is the rule, not the exception.

\subsection{Occupancy cap: registers and shared memory}\label{sec:rs-bind}

\Cref{sec:warp-pool} shows that the warp-pool cap is set by the
\emph{kernel}, not by the architecture.  The two resources that most
often bind are the register file and shared memory:
\begin{align*}
R_\text{eff}     &= \lceil R / 8 \rceil \cdot 8, \\
\eta_\text{reg}  &= \min\!\left(64,\ W_\text{block} \cdot \left\lfloor \tfrac{65536}{R_\text{eff} \cdot W_\text{block} \cdot 32} \right\rfloor\right), \\
\eta_\text{smem} &= \min\!\left(64,\ \left\lfloor \tfrac{228 \cdot 1024}{S_\text{block}} \right\rfloor \cdot W_\text{block}\right),
\end{align*}
with binding cap \(\min(\eta_\text{reg}, \eta_\text{smem}, 64)\).
Here \(R\) is NCU-reported registers per thread and \(R_\text{eff}\)
rounds \(R\) to Hopper's 8-register/thread allocation granularity;
\(S_\text{block}\) is shared memory per block and \(W_\text{block}\)
is warps per block.  Per-kernel values are in \Cref{tab:rscaps}.

The \texttt{cuBLASLt} rows use the tile that NVIDIA's nvjet heuristic
selects for all four dense projection roles on all four models
(Llama-3-8B decode shown; \role{o\_proj} and \role{down\_proj} land on
the same 12-warp cap as \role{qkv} and Gate/Up and are omitted as
duplicates).  The CUTLASS rows are the three autotuned tiles within a
single FP8 kernel (\texttt{cutlass\_3x\_gemm\_sm90\_fp8}), covering 15
distinct tile configurations across the four models.
The \emph{Bind.} column names the binding resource: ``both'' when
$\eta_\text{reg} = \eta_\text{smem}$ (co-limited), ``smem'' when SMEM
alone binds.
\begin{table}[ht]
\centering
\caption{Per-resource warps/SM ceilings, grouped by kernel
family.
%The cuBLASLt rows use the tile that NVIDIA's nvjet heuristic selects for all four dense projection roles on all four models (LLaMA-3-8B decode shown; \role{o\_proj} and \role{down\_proj} land on the same 12-warp cap as \role{qkv} and Gate/Up and are omitted as duplicates).  The CUTLASS rows are the three autotuned tiles within a single FP8 kernel (\texttt{cutlass\_3x\_gemm\_sm90\_fp8}), covering 15 distinct tile-config observations across the four models.  The \emph{Bind.} column names the binding resource: ``both'' when $\eta_\text{reg} = \eta_\text{smem}$ (co-limited), ``smem'' when SMEM alone binds.
}
\label{tab:rscaps}
\footnotesize
\setlength{\tabcolsep}{3pt}
\begin{tabular}{@{}lrrrrrrl@{}}
\toprule
Configuration & Block & R/thr & SMEM & \(\eta_\text{reg}\) & \(\eta_\text{smem}\) & Cap & Bind. \\
\midrule
\multicolumn{8}{@{}l}{\emph{\texttt{cuBLASLt} nvjet (BF16) --- 4 projection roles}}\\
QKV proj.\           & 384 & 168 & 160\,KB & 12 & 12 & 12 & both \\
Gate/Up proj.\       & 384 & 168 & 220\,KB & 12 & 12 & 12 & both \\
\midrule
\multicolumn{8}{@{}l}{\emph{\gls{fa3} (BF16) --- decode path}}\\
\gls{fa3} main        & 256 & 255 & 147\,KB &  8 &  8 &  8 & both \\
\midrule
\multicolumn{8}{@{}l}{\emph{CUTLASS \texttt{sm90\_fp8} (FP8) --- 3 autotuned tiles}}\\
Small tile           & 256 &  34 & 227\,KB & 48 &  8 &  8 & smem \\
Medium tile          & 256 &  58 & 198\,KB & 32 &  8 &  8 & smem \\
Large tile           & 384 & 168 & 215\,KB & 12 & 12 & 12 & both \\
\bottomrule
\end{tabular}
\end{table}

\gls{fa3} main's 256-thread block (8~warps, \Cref{tab:rscaps}) reflects the
\texttt{use\_one\_mma\_wg} decode heuristic in vLLM-\gls{fa3}:
for $\text{seqlen\_q} \le 64$ on sm90, the kernel selects a smaller
tile (kBlockM\,=\,64) that allocates 256 registers per MMA thread,
driving $\eta_\text{reg}$ and $\eta_\text{smem}$ both to 8~warps/SM ---
half the dense-GEMM cap of~12.

The dense BF16 GEMM tile configuration is selected by
\texttt{cuBLASLt}'s nvjet heuristic, whose source is proprietary;
therefore, the BF16 rows in \Cref{tab:rscaps} are verified via NCU's
runtime-reported occupancy-limiter counters rather than source
inspection.

Surprisingly, every dense BF16 kernel we profiled is
\emph{co-limited}: $\eta_\text{reg}$ and $\eta_\text{smem}$ land on
the \emph{same} warp count, so neither resource is the sole
limiter.  Across projection roles, models, and regimes, the cap is
uniformly 12~warps/SM with $\eta_\text{reg}{=}\eta_\text{smem}$; the
underlying \texttt{cuBLASLt} nvjet tiles reduce to a small set of
distinct (regs, SMEM, block) configurations, every one balanced.
The cap is thus a property of the tiles \texttt{cuBLASLt}'s nvjet
heuristic selects for BF16, not of context length, batch size, or
Hopper itself: changing $(S, B)$ does not change it.

A separate FP8-E4M3 sweep (the \texttt{cutlass\_3x\_gemm\_sm90\_fp8}
path; 4 models, prefill and decode; 11.3\,k kernel launches)
makes this concrete.  The three dominant tile configurations
(\Cref{tab:rscaps}, lower rows) fall into two groups:
\emph{(i)}~two SMEM-only-bound small/medium configurations
(34--58 regs/thread, $4\text{--}6\times$ register headroom)
covering $\approx$90\% of instances, and
\emph{(ii)}~one register/SMEM co-limited large configuration
(168 regs/thread) covering the remaining $\approx$10\%.
The first group caps the kernel at 8 warps/SM through shared
memory alone; the second recovers the BF16-style 12/12 balance.
The lever is therefore tile- and precision-specific: a one-sided
SMEM relaxation deepens the pool only for the SMEM-bound FP8
configs (8$\to$16 warps/SM), never for the co-limited BF16 GEMMs.
(FP8 is otherwise out of scope; CUTLASS hand-tuned kernels and
decode-attention paths outside \texttt{cuBLASLt} remain untested.)
\FloatBarrier
\subsection{Stall-reason decomposition}\label{sec:stalls}

Given the warps a kernel keeps resident, the next question is why they fail to issue every cycle. We use NCU's per-issue stall breakdown of \texttt{warp\_cycles\_per\_issued}, which we call the \emph{stall budget}: average warp stall cycles per issued instruction, grouped by stall reason.

\begin{figure}[t]
\centering
\begin{subfigure}[t]{0.49\linewidth}
  \centering
  \includegraphics[width=\linewidth]{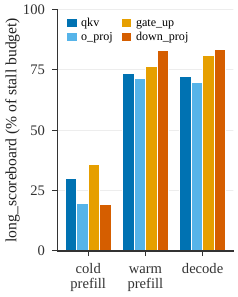}
  \caption{Memory-dependency stalls.}
  \label{fig:stall-wave-a}
\end{subfigure}
\hfill
\begin{subfigure}[t]{0.49\linewidth}
  \centering
  \includegraphics[width=\linewidth]{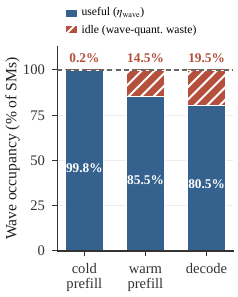}
  \caption{Launch-grid wave quant.}
  \label{fig:stall-wave-b}
\end{subfigure}
\caption{Two regime-resolved metrics for dense BF16 GEMMs, cell-equal-weight mean over the H100 NCU sweep. \textbf{(a)} \texttt{long\_scoreboard} share of the per-cycle stall budget. \textbf{(b)} Wave efficiency $\eta_\text{wave}$ (solid) and wave-quantization waste ($100-\eta_\text{wave}$, hatched: SMs idle in the trailing wave), stacked to 100\%.}
\label{fig:stall-wave}
\end{figure}

NCU reports seventeen stall reasons; for the dense BF16 GEMMs in our sweep, two dominate. \texttt{long\_scoreboard} indicates a warp waiting on a long-latency memory load (bandwidth-bound), while \texttt{math\_pipe\_throttle} indicates that the compute pipeline is the bottleneck (compute-bound). \Cref{fig:stall-wave-a} shows the \texttt{long\_scoreboard} share by regime. The companion panel (\Cref{fig:stall-wave-b}) reports the wave-quantization metric from \S\ref{sec:wave}; we show them together because they move in \emph{opposite} directions across regimes.

The pattern is clean.  Decode and warm prefill are dominated by
memory-dependency stalls: \texttt{long\_scoreboard} accounts for
70--84\% of the per-cycle stall budget, consistent with
bandwidth-bound execution and substantial headroom in the issue
pipeline.  Cold prefill drops to 19--36\%: the larger
\texttt{cuBLASLt} tiles amortize the same loads across more compute,
and the remaining stall budget shifts to \texttt{gmma},
\texttt{barrier}, and \texttt{mio\_throttle}.

\texttt{math\_pipe\_throttle} does not appear in the top-3 stall
reasons for any of the 360 dense-GEMM rows (182 cold-prefill, 42
warm-prefill, 136 decode; one per (model,\,role,\,$S$,\,$B$) tuple,
all with complete stall counters; \gls{fa3}/MoE/aux rows excluded).
Accordingly, ``compute-saturated'' is not the right description even
for the busiest cold-prefill cells, which are issue-pipeline
limited via \texttt{gmma} barriers, not pipeline-throttled.

\gls{fa3} diverges from this picture as its dominant stalls are
\texttt{stall\_wait}/\texttt{barrier}, not \texttt{long\_scoreboard}.
\S\ref{sec:b32} quantifies this contrast and its memory-fix
implication.

\begin{figure}[t]
\centering
\includegraphics[width=0.9\linewidth]{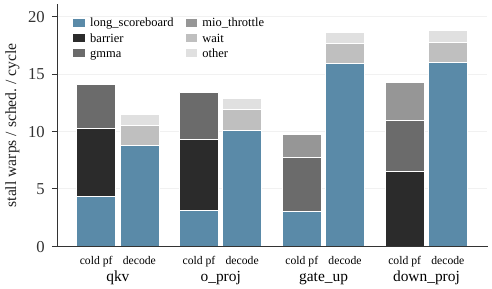}
\caption{Stall-reason breakdown of \texttt{warp\_cycles\_per\_issued} for Llama-3-8B dense BF16 GEMMs (cold prefill (P) vs. decode (D)) at $(S{=}2048, B{=}1)$. Cold prefill is barrier/\texttt{gmma}-dominated, while decode is dominated by \texttt{long\_scoreboard}. \emph{other} aggregates NCU \texttt{selected} (issue) cycles and remaining minor stalls.%
% This pattern is stable across the sweep: decode \texttt{long\_scoreboard} stays 64--81\% and cold prefill remains barrier/\texttt{gmma}-dominated for all $(S,B)\in\{2\text{k}\dots8\text{k}\}\times\{1\dots8\}$. 
}
\label{fig:stalls}
\end{figure}

Concretely, Llama-3-8B QKV at \((S{=}2048, B{=}1)\) is
\texttt{long\_scoreboard}-dominated and effectively identical in
decode and warm prefill --- 72.7\% and 72.8\% of
\texttt{warp\_cycles\_per\_issued}, respectively.  In cold prefill at
$(S{=}8190, B{=}4)$ the same kernel runs the longer \texttt{cuBLASLt}
tile at 20.85 cycles per issued instruction, of which only 23.9\% are
\texttt{long\_scoreboard}; the largest contributors are
\texttt{barrier} (27.6\%) and \texttt{gmma} (23.3\%), the remainder
\texttt{mio\_throttle}, \texttt{wait}, and the issue slot.  The issue
pipeline saturates on the GMMA pipeline and inter-warp
synchronization, \textbf{not} memory. \Cref{fig:stalls} plots the full breakdown; the per-reason values
are concrete fix-it targets for code-generation teams.

\subsection{Matmul throughput against the whole array}\label{sec:devsol}

The per-active-SM and warp-pool metrics quantify how hard the SMs a
kernel \emph{occupies} are working, but \textbf{not} how much of the
H100's aggregate matmul capacity is exercised over the kernel's
wall-clock.  To capture whole-device matmul throughput, we take
\gls{ncu}'s \texttt{compute\_sol\_pct} (the \emph{Compute (SM)
Throughput} SOL metric on the SMs the kernel reaches, \S\ref{sec:method})
and scale it by spatial coverage to obtain \emph{device-wide compute
SOL}, a per-kernel analog of model-level \gls{mfu}:
\[
\text{device-wide compute SOL} = \text{compute\_sol\_pct} \cdot \tfrac{\text{SMs\_reached}}{132}.
\]

This scaling accounts for spatial \emph{reach} only; it does \emph{not}
discount multi-wave tail loss (the wave-efficiency metric of
\S\ref{sec:wave}), so for multi-wave launches it is an upper bound on
true elapsed whole-device duty.  We also report ``effective active SM
compute throughput'' ($528 \times$ device-wide compute SOL) as a
wall-clock-duty upper bound, not a literal Tensor-Core count.

\smartparagraph{The cross-counter triangulation.}  The device-wide
compute SOL is corroborated by two \emph{independent} NCU counter
families read on the \emph{same kernel call}: the DRAM-throughput SOL
(\texttt{gpu\_\_dram\_throughput.}\allowbreak\texttt{avg.}\allowbreak\texttt{pct\_of\_peak\_sustained\_elapsed})
and the stall budget (\S\ref{sec:stalls}).
On the dense decode GEMM mean the device-wide compute SOL is
$\sim$7.9\% while the DRAM SOL is $\sim$64\% (both elapsed-cycle
denominator); the $\sim$8$\times$ asymmetry
falsifies ``the kernel is idle'' without invoking stall
classifiers.  The stall-budget view (\S\ref{sec:stalls},
\texttt{long\_scoreboard} 70--84\%) is the second corroborating
counter family: it is sourced from the per-issue warp-stall counters,
independent of \texttt{sm\_\_throughput} and of DRAM throughput.

\subsection{The GMMA fragment-fill ceiling on decode matmul}\label{sec:tc}

Independent of stalls and spatial coverage, a
\emph{kernel-internal} ceiling caps useful matmul on small-$M$
shapes.  Hopper's \gls{bf16} \gls{gmma} instruction issues at a
fixed $m64{\times}n\{8,\ldots,256\}{\times}k16$
fragment~\cite{wgmma}; any GEMM with $M_\text{actual}{<}64$ still
issues full $m64$ instructions and pads the unused rows with
zeros.  We define the $M$-axis fragment fill
$\eta_M = M_\text{actual} / (\lceil M_\text{actual}/64 \rceil \cdot 64)$.
Across the 360-row dense-GEMM grid (the dense-role subset of the
\Cref{tab:coverage} rows; $\text{4 models}\times\text{4 roles}\times S\times B$):
$\eta_M{=}100\%$ in all 182 cold-prefill rows ($M{=}S{\cdot}B{\gg}64$);
$\eta_M{=}25{-}100\%$ in warm prefill (42 rows; $M{\approx}16B$, mean 70.2\%);%
% (From the 16-token cache block described in \S\ref{sec:regimes}.)
and $\eta_M{=}1.56,\,3.12,\,6.25,\,12.5\%$ in decode
for $B{\in}\{1,2,4,8\}$ (136 rows, mean 5.86\%), with
\emph{no} exceptions across the four models or four dense roles.
This is the kernel-internal complement of the spatial wave-quantization
of \S\ref{sec:wave}: wave efficiency charges idle SMs in the
trailing wave; $\eta_M$ charges idle MMA-rows within each
warp-group instruction.  Both compound against the
device matmul-peak budget.

The consequence is a metric-interpretation failure mode in
\texttt{compute\_sol\_pct}.  On Llama-3-8B decode \role{qkv} at
$(S{=}2048, B{=}1)$, \texttt{compute\_sol\_pct}~$=9.5\%$ per
active SM, reaching 96 of 132 SMs; device-wide compute SOL is
$9.5\%{\times}96/132{=}6.9\%$.  But with $\eta_M{=}1.56\%$, the
useful FLOP throughput against the 132-SM 835~TFLOP/s peak is
$9.5\%{\times}1.56\%{\times}96/132{\approx}0.11\%$ (${\approx}0.9$~TFLOP/s),
because the SM-throughput counter charges padded $m64$ MMAs as work
(for a dense BF16 GEMM the tensor pipe is the throughput-limiting
sub-pipeline) and cannot distinguish them from useful rows.  This is a
metric-interpretation claim, not a recoverable-headroom claim: the
$M{=}1, N{=}6144, K{=}4096$ shape is bandwidth-bound and no
\texttt{cuBLASLt} tile selection ($\text{BLOCK}_M \geq 64$ is
hardware-bound; \S\ref{sec:wave}) can change this.  The $6.9\%$
reading therefore overstates useful matmul by
${\sim}1/\eta_M{\approx}64{\times}$ rather than indicating that
${\sim}90\%$ of the SM compute throughput is recoverable through
better tiling.  The lever
(\S\ref{sec:implications}) is to raise $M$ itself, via
persistent-decode kernels or cross-request token packing, not to
retile.\footnote{Throughout, $\eta_M$ refers to the
$M$-axis only; the \gls{gmma} $N$-axis ranges 8--256 and is
\texttt{cuBLASLt}-selected, so $N$ is not similarly wasted.}

\begin{figure}[t]
\centering
\includegraphics[width=0.9\linewidth]{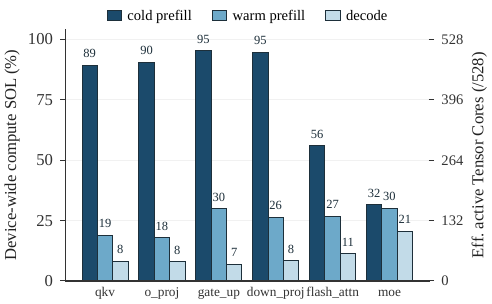}
\caption{Device-wide compute SOL on H100 by regime (\nm{compute\_sol\_pct}$\times$SMs reached/132). Right axis: effective SM compute-throughput equivalents (out of 528), an intuition-only upper bound, not a literal Tensor-Core count. Averaged over the four dense GEMMs, cold prefill reaches 92.0\% device-wide SOL but decode only 7.9\%, an 11.7$\times$ gap.}
\label{fig:tc-headline}
\end{figure}

Small-batch dense decode delivers a low fraction of the chip's
matmul throughput in \emph{absolute} terms, not merely relative to
prefill: the device-wide compute SOL means by regime
(\Cref{fig:tc-headline}) put decode at 7.9\% of peak against
92.0\% for cold prefill and 22.6\% for warm prefill, even though
decode launch grids reach 96--132 of 132 SMs.  The warm figure is low
for a different reason, since the \texttt{cuBLASLt} cache has traded
the cold tile for an essentially decode-sized one.  Per-active-SM
compute SOL is the \emph{same} \gls{ncu} \emph{Compute (SM) Throughput}
counter as \smbusy{} --- both are
\texttt{sm\_\_throughput.}\allowbreak\texttt{avg.}\allowbreak\texttt{pct\_of\_peak\_sustained\_elapsed},
column-identical across all 759 profiled kernels (max difference
0.16\,pp).  It differs from the device-wide form only by the
$\text{SMs\_reached}/132$ factor of \S\ref{sec:devsol}, so it reads higher
but tells the same cold/\allowbreak warm/\allowbreak decode story: the
dense-GEMM mean is 92\,/\,25\,/\,9\% (\Cref{tab:cross-metric}), with
\role{flash\_attn} at 55.9\,/\,26.9\,/\,11.3\% and the MoE-experts
kernel at 31.6\,/\,30.0\,/\,20.5\%.

\subsection{Wave-quantization waste}\label{sec:wave}

The fragment floor of \S\ref{sec:tc} wastes rows inside each instruction.
The launch grid can waste whole SMs in the same way, one level up.
A CUDA grid with \(N_\text{CTA}\) thread blocks (CTAs) executes on 132
SMs in \(\lceil N_\text{CTA}/132 \rceil\) waves of up to 132 blocks
per wave.  The final wave is full only if \(132 \mid N_\text{CTA}\);
otherwise, \(132 - (N_\text{CTA} \bmod 132)\) SMs sit idle for the
entire duration of the trailing wave, yet the kernel is still charged
that wave's wall-clock.  Useful array utilization is therefore
\emph{quantized} to 132-CTA multiples and rounded up to the next full
wave (NVIDIA's tail-/wave-quantization effect).

We quantify this effect with \emph{wave efficiency} $\eta_\text{wave}$,
the fraction of per-wave SM slots, summed across all waves, that do
useful work:
\(\eta_\text{wave} = N_\text{CTA} / (\lceil N_\text{CTA}/132 \rceil \cdot 132)\); and report per-call wasted SM-slot time,
\(T_\text{wasted} = (1 - \eta_\text{wave}) \cdot T_\text{kernel} \cdot 132\)
SM-\(\mu\)s (idle SM capacity, not elapsed wall-clock).
\Cref{fig:stall-wave-b} decomposes the per-regime SM-wave budget
into useful $\eta_\text{wave}$ and wave-quantization waste
($100-\eta_\text{wave}$): 0.2\% in cold prefill, 14.5\% in warm
prefill, and 19.5\% in decode (dense GEMMs, cell-equal-weight mean
over the NCU sweep).  Worst single calls waste $\sim$21\,K SM-$\mu$s.

\begin{figure}[t]
\centering
\includegraphics[width=0.9\linewidth]{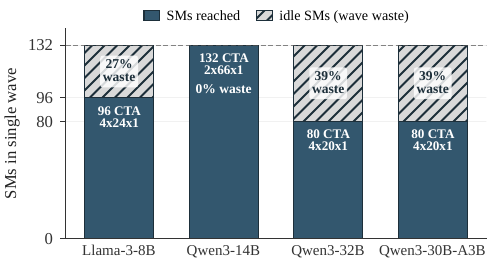}
\caption{Decode-QKV grid CTA count per model: 132-CTA (Qwen3-14B,
0\% waste), 96-CTA (Llama-3-8B, 27\%), 80-CTA (Qwen3-32B and
-30B-A3B, 39\% each). Grid is invariant across all swept
$S{\in}\{2048,\dots,16384\}$, $B{\in}\{1,2,4,8\}$.}
\label{fig:wave}
\end{figure}

The pathology is sharp but model-specific (\Cref{fig:wave}).
Llama-3-8B QKV launches 96 CTAs ($\eta{=}72.7\%$, 36 idle SMs,
$\sim$21\,K SM-$\mu$s wasted per call).  Qwen3-32B and Qwen3-30B-A3B
launch 80 CTAs ($\eta{=}60.6\%$, 39\% waste).  Qwen3-14B happens
to fill the chip at 132 CTAs.  Other Llama dense projections use
$N_\text{CTA}{\in}\{128,132\}$ ($\eta{=}97$--$100\%$).  \gls{fa3} is
100\% wave-efficient (persistent kernel, internal tile loop).
Wave efficiency is the \emph{spatial} complement of the \gls{gmma}
$M$-fragment fill $\eta_M$ (\S\ref{sec:tc}); the
$\text{BLOCK}_M{\geq}64$ floor that bounds $\eta_M$ also bounds the
tile-space sweep below (no producible \texttt{cuBLASLt} tile has
$\text{BLOCK}_M{<}64$).

\begin{figure}[t]
\centering
\includegraphics[width=0.95\linewidth]{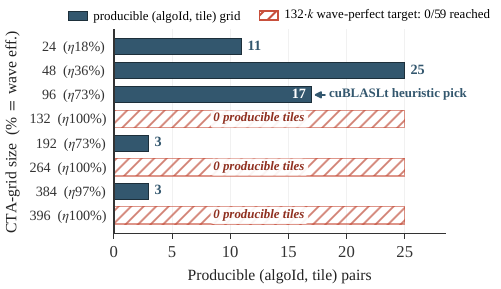}
\caption{\texttt{cuBLASLt} tile-space enumeration for Llama-3-8B decode-QKV ($M{=}8$, $N{=}6144$, $K{=}4096$, BF16 NT) on H100. Bars count producible (\textit{algoId}, \textit{tile}) pairs at each resulting CTA-grid size; $\eta$ is the corresponding wave efficiency. The three 132$\cdot k$ wave-perfect grids are interleaved as hatched rows: \emph{none} of the 59 producible pairs reaches one, and the heuristic settles on 96 CTAs.}
\label{fig:tile-ab}
\end{figure}

We then ask whether a 132-multiple tile exists within
\texttt{cuBLASLt}'s algorithm space.  We swept the full
$17\times12=204$-entry space of \texttt{cuBLASLt}'s 17 BF16 NT
\hbox{algoIds} $\times$ 12 tile shapes for the Llama-3-8B decode-QKV problem
($N{=}6144, K{=}4096$) at the $M$ where \texttt{cuBLASLt} selects the
96-CTA tile ($M{=}8$; the heuristic holds the same tile across
$M{\in}\{8,16,32,64,128\}$).  Of the 17 algoIds, 10 expose no usable
tile for this shape, leaving \textbf{59}
(\textit{algoId}, \textit{tile}) pairs.  The finding is clearly
\emph{negative}: \emph{zero} of the 59 pairs yields a 132-multiple
grid for this shape (\Cref{fig:tile-ab}), every producible grid
falling in $\{24, 48, 96, 192, 384\}$.

The reason is structural: since $132 = 4{\cdot}3{\cdot}11$ contains a
factor 11 and QKV has $N{=}6144 = 2^{11}{\cdot}3$, no power-of-2 tile
shape can produce a CTA grid size
$\lceil M/M_t \rceil \cdot \lceil N/N_t \rceil \in \{132, 264, \dots\}$
for these dimensions.  A 132-multiple grid would need either a
non-standard tile factor (e.g., a 559--614-wide $N$-tile, giving
exactly 11 $N$-tiles) or split-K with factor 11, neither present in
\texttt{cuBLASLt}'s current nvjet tile set for this shape.  The 96-CTA tile
\texttt{cuBLASLt} selects is nonetheless defensible on per-CTA
efficiency grounds: enumerated alternatives that fill the chip better
(e.g., $16{\times}16$ at 384 CTAs, 97\% wave efficiency) shrink
per-CTA work substantially, and \texttt{cuBLASLt}'s heuristic
optimizes throughput rather than wave fill alone.

\subsection{Memory hierarchy and L2 working-set fit}\label{sec:l2}

Decode is HBM-bandwidth bound (\S\ref{sec:stalls}), but separating
kernels that are limited by L2 capacity from those limited by HBM
bandwidth is a finer question.  We use the L2 fit ratio
$\rho = \text{footprint}/60\,\text{MB}$, where $\rho>1$ indicates that
the working set spills L2 across calls.

Decode FFN projections (\texttt{gate\_up}, \texttt{down\_proj}) spill
L2 by 2--4$\times$ across all dense models, consistent with the low
observed L2 hit rates.  Decode QKV is dominated by its weight matrix
($\sim$48\,MB on Llama-3-8B), so fits L2 ($\rho{\approx}0.80$),
whereas the larger Qwen QKV weights spill (1.2--1.7$\times$).

Across all (kernel, $S$, $B$) combinations, 90.0\% spill in cold
prefill, 92.5\% in warm prefill, and 53.8\% in decode.
The per-call MoE \texttt{moe} row spills less aggressively than dense
FFNs at $B{=}1$: top-8 of 128 experts touches roughly $1/16$ the
dense-FFN working set, so the per-call footprint is correspondingly
smaller.  At $B{>}1$, up to $\min(Bk,E)$ distinct expert weight
matrices may be loaded per call, raising the effective MoE weight
footprint by up to $\min(B,\,16){\times}$; thus, the $\rho{\approx}1.20$
figure is a lower bound at larger batches (see the working-set caveat
in \S\ref{sec:threats}).

\subsection{Compute-bound crossover batch}\label{sec:ridge}

Relative to the BF16 roofline ridge $I^{*}\!\approx\!214$ FLOP/byte
(\S\ref{sec:bg}), we
capture where each role lives with the compute-to-memory SOL ratio
$r(B)=\text{compute\_sol\_pct}(B)/\text{memory\_sol\_pct}(B)$, and
call the smallest $B$ for which $r(B)\!\geq\!1$ --- a SOL-balance
proxy for the roofline ridge (achieved tensor- vs.\ memory-pipe
duty, \textbf{not} an arithmetic-intensity computation) --- the
\emph{crossover batch} $B^{*}$. Within the swept range
($B\!\leq\!8$), no decode dense-GEMM crosses for any model. Per-role
$r$ values are uniformly low and tightly clustered for the dense
projections (qkv 0.21, o\_proj 0.26, gate\_up 0.12, down\_proj 0.13);
flash\_attn alone is closer at 0.84. Linear extrapolation puts
$B^{*}$ at $\sim$31, $\sim$25, $\sim$49, $\sim$45, and $\sim$8
respectively.
Wave quantization (\S\ref{sec:wave}) therefore dominates well before
$B^{*}$ is reached. Production decode
rarely runs $B{\geq}25$ per request because of the latency \gls{slo}, so
dense decode GEMMs stay memory-bound.

\subsection{Empirical validation at \texorpdfstring{$B{=}32$}{B=32}}
\label{sec:b32}

The crossover batches in \S\ref{sec:ridge} are linear
extrapolations beyond the swept \(B\!\leq\!8\) range; without a
measurement at \(B\!\geq\!25\) the prediction is unfalsifiable.
We therefore extend the sweep with one $B{=}32$ validation H100/\gls{fa3} NCU
pass per model at the largest \(B\) the KV-memory budget admits:
\(B\!=\!32\) at \(S\!=\!8190\) on Llama-3-8B, \(S\!=\!16384\) on
Qwen3-14B, and \(S\!=\!8192\) on the sparse-MoE Qwen3-30B-A3B; the
62\,GB Qwen3-32B caps at \(B\!=\!4\) at \(S\!=\!16384\) under the
same 0.92 \texttt{gpu\_memory\_utilization}, and so appears twice in
\Cref{tab:b32} (\(B{=}4\) at long \(S\), \(B{=}32\) at short \(S\)).
Every other lever (vLLM~0.16, \gls{fa3}, BF16, prefix-caching ON,
\(\geq\)3 warmups) is held identical to the headline sweep.
\Cref{tab:b32} reports before/after comparisons: each \(B{=}32\)
validation row (top panel; five rows) against the same model's
\(B{=}1\) reference (bottom panel), at matched context length where
the KV budget allows and cross-\(S\) otherwise (footnoted).

\begin{table}[t]
\centering
\caption{Decode microarchitecture at \(B{=}32\) on H100 with \gls{fa3}. Columns report per-iteration time shares for the three largest buckets (dense GEMMs / \gls{fa3} / MoE experts), SM-busy for dense GEMMs, and per-token decode time (ms/tok). Shares are not a partition and may not sum to 100\%.}
\label{tab:b32}
\setlength{\tabcolsep}{1pt}
\scriptsize
\begin{tabular}{@{}llrrrrr@{}}
\toprule
Model & ($S$, $B$) & dense\% & \gls{fa3}\% & moe\% & SM-busy & ms/tok \\
\midrule
\multicolumn{7}{l}{\emph{\(B{=}32\) validation arm ($B$ at memory cap):}} \\
Llama-3-8B$^\S$    & (8190, 32)  & 58.2 & 32.7 & --- & 51.7 & 1.52 \\
Qwen3-14B     & (16384, 32) & 44.0 & \textbf{49.5} & --- & \textbf{71.9} & 2.38 \\
Qwen3-30B-A3B & (8192,  32) & 11.4 & 31.6 & 36.2 & 34.4 & 0.96 \\
Qwen3-32B     & (16384,  4) & 69.6 & 18.8 & --- & 23.8 & 18.4 \\
Qwen3-32B     & (2048,  32) & 79.2 & 11.6 & --- & \textbf{72.6} & 0.98 \\
\midrule
\multicolumn{7}{l}{\emph{$B{=}1$ reference (matched $S$ except where footnoted):}} \\
Llama-3-8B$^\S$    & (8190,  1)  & 80.5 &  9.8 & --- & ${\sim}10$ & 5.50 \\
Qwen3-14B     & (16384, 1)  & 80.8 & 10.5 & --- & ${\sim}12$ & 10.10 \\
Qwen3-30B-A3B & (16384, 1)$^\dagger$ & 21.8 & 15.2 & 35.7 & ${\sim}\,9$ & 5.32 \\
Qwen3-32B     & (16384, 1)$^\ddagger$  & 85.8 &  7.6 & --- & ${\sim}\,8$ & 22.91 \\
\bottomrule
\multicolumn{7}{@{}l}{\scriptsize $^\dagger$$B{=}1$ ref reported at $S{=}16$K (long-context regime); the}\\
\multicolumn{7}{@{}l}{\scriptsize \;\;\;cross-$S$ comparison to the $(8$K$,32)$ cell is sound because dense}\\
\multicolumn{7}{@{}l}{\scriptsize \;\;\;SM-busy is $S$-invariant ($8.9$--$9.0$\% at all swept $S$, incl.\ $8$K).}\\
\multicolumn{7}{l}{\scriptsize $^\ddagger$shares at $S{=}16$K; ms/tok column is at $S{=}32$K (the available cell).}\\
\multicolumn{7}{@{}l}{\scriptsize $^\S$\texttt{meta-llama/\allowbreak Meta-Llama-3-8B} has \texttt{max\_position\_\allowbreak embeddings}$=$$8192$;}\\
\multicolumn{7}{@{}l}{\scriptsize \;\;\;$S{=}8190$ leaves 2-token decode headroom under that $8192$ limit}\\
\multicolumn{7}{@{}l}{\scriptsize \;\;\;($8190{+}2{=}8192$), not a measurement choice.}\\
\end{tabular}
\end{table}

\smartparagraph{Crossover and bottleneck shift at \(B{=}32\).}
On Qwen3-14B, decode dense-GEMM SM-busy rises from \(\sim\)12\%
at \(B{=}1\) to \(\mathbf{71.9\%}\) at \(B{=}32\); \glsentryshort{fa3}-fwd
reaches \(\mathbf{49.5\%}\) of decode iteration, overtaking
dense GEMMs (44.0\%); per-token wall-clock falls from 10.10 to
2.38\,ms ($\sim$4.2$\times$); \texttt{long\_scoreboard} on dense
GEMMs drops from 26 to 12 cycles/issue and DRAM busy from 39 to
36\% --- the measured cell lands near the predicted
\(B^{*}{\approx}31\) for QKV.
Llama-3-8B approaches but does not cross: SM-busy 51.7\%
(\(5\times\) the \(B{=}1\) reference), \gls{fa3} 32.7\%, dense still
58.2\%; its smaller hidden \(F{=}14336\) (vs.\ 17408) places
\(B^{*}\) at \(\sim\)49 for \role{gate\_up}, beyond a feasible
SLO batch.
On Qwen3-30B-A3B the bottleneck shifts: the routed-FFN share of decode time
(the \fusedmoe{}\% column of \Cref{tab:b32}: \fusedmoe{} plus its
align/sum/scatter aux, router excluded) is essentially flat from
35.7\% at \(B\!=\!1, S\!=\!16\)\,K to 36.2\% at
\(B\!=\!32, S\!=\!8\)\,K, while \glsentryshort{fa3}-fwd \emph{more than doubles}
from 15.2\% to 31.6\%.  At \(B\!=\!32\) the model is jointly
routed-FFN and \gls{fa3} bound rather than routed-FFN dominated, and
the two share decode time within 5\% points.

Qwen3-32B crosses at short \(S\) but is KV-forbidden at long \(S\):
at the long-context cell
(\(B{=}4, S{=}16\)K) dense SM-busy reaches 23.8\% --- above the
small-batch mean of 8.2\% but far from \(B^{*}{\approx}49\).
At (\(B{=}32, S{=}2048\)) the same kernels reach \textbf{72.6\%}
SM-busy --- matching Qwen3-14B's \(71.9\%\) at (16K, 32),
i.e.\ the dense stream crosses \emph{despite} the 32B parameter
count --- and per-token wall-clock collapses to 0.98\,ms, on par
with the sparse MoE's 0.96\,ms at (8K, 32) even though the MoE
activates only ${\sim}3$B params/token.  The crossover is a \(B\) effect on the
dense stream; the KV footprint (KV bytes/token $= 2{\cdot}n_\text{kv}{\cdot}d_\text{head}{\cdot}2{\cdot}L$:
256\,KB for Qwen3-32B with $n_\text{kv}{=}8, L{=}64$, vs.\ 96\,KB
for the MoE with $4, 48$) constrains \emph{when}, not
\emph{whether}: \(B{=}32\) needs \(S{\lesssim}3\)K on the 94\,GB
part, while long-\(S\) (16\,K) loads are KV-capped to \(B{\leq}4\)
in our sweep (\S\ref{sec:workload}).  This refines
the right-batching implication (\S\ref{sec:implications}): shifting
the operating-point question from ``can model \(M\) cross?'' to
``does \((S, B)\) admit it?''

\gls{fa3}'s stall signature stays distinct from dense decode at
\(B{=}32\): dense GEMMs keep the canonical
\texttt{long\_scoreboard} (memory-dependency) signature
(11.5--26.4 cycles/issue), but \glsentryshort{fa3}-fwd's top stall is
\texttt{stall\_wait} (barrier/fixed-latency dependency, 1.8) with
\texttt{long\_scoreboard} only 1.1--1.2: HBM upgrades move dense
GEMMs but not \gls{fa3}, and the \glsentryshort{fa3} share growing to \(\sim\)50\% of
decode reflects attention compute scaling with \(B\), not a hidden
bandwidth bottleneck.  This sharpens the \glsentryshort{fa3}-rework implication
(\S\ref{sec:implications}): \(B{=}32\) \hbox{decode} latency needs
kernel-side rework on \glsentryshort{fa3}, distinct from the memory-hierarchy
levers that address the dense stream.

Finally, KV-cache write closes the cost composition at
\(<\!2\%\) even at \(B{=}32\): the
\texttt{reshape\_and\_cache\_flash} kernel stays at 0.7--1.6\%
of decode iteration across all four \(B\!=\!32\) cells (Llama-3-8B
1.19\%, Qwen3-14B 0.69\%, Qwen3-30B-A3B 1.55\%, Qwen3-32B
1.26\%).  The atomic decomposition used by
chunked-prefill / PD-disaggregation cost
models~\cite{sarathi,distserve,splitwise} ---
\(T_\text{iter}{\approx}T_\text{dense}{+}T_\text{\glsentryshort{fa3}}\) for dense
models, \(T_\text{iter}{\approx}T_\text{dense}{+}T_\text{\glsentryshort{fa3}}{+}T_\text{moe}\)
for the MoE --- therefore needs no separate \(T_\text{KV}\) term:
the KV-write residual stays \(\leq\!2\%\) through \(B{=}32\).

\subsection{Kernel-selection stability}\label{sec:dispatch}

Stable kernel selection is a precondition for hand-tuning and
persistent-kernel rewrites.  We measure \emph{kernel-selection
diversity} as the number of distinct kernel names that
\texttt{cuBLASLt} selects for each (model, regime, role) across the
sweep.  In decode, this count is uniformly 1 for dense GEMMs and 2
for \gls{fa3}; cold-prefill QKV peaks at 7 (Llama-3-8B).  Thus, decode
underperformance is not caused by unstable kernel selection:
\texttt{cuBLASLt} selects kernels stably, and the selected kernels
remain \emph{inefficient} --- a favorable setting for
persistent-kernel rewrites.

\begin{table*}[t!]
\centering
\caption{Cross-metric, cross-regime summary on dense BF16 GEMMs.
The metrics split into \emph{code-generation levers} (warps/SM, wave
efficiency --- largely flat across regimes) and \emph{workload
outcomes} (\smbusy{}, compute SOL, L2 fit, launch overhead, stalls
--- order-of-magnitude shifts cold$\to$decode); this separates fixes
that live in \texttt{cuBLASLt}/compilers/serving from those that
require workload restructuring.  The launch-overhead row is
analytical ($\tau{=}5\,\mu$s).}
\label{tab:cross-metric}
\begin{tabular}{lllll}
\toprule
Metric & Cold prefill & Warm prefill & Decode & Cold$\to$decode \\
\midrule
SM time-utilization (\smbusy)         & 73--97\% & 8--73\% & 6.5--12.7\% & $\sim$10$\times$ (mean 92\% vs.\ 9.4\%) \\
Achieved warps/SM                     & 8.9--9.5 & 8.8--9.5 & 8.8--9.5 & $\approx 1\times$ (flat) \\
Compute SOL, per active SM (dense GEMM mean) & 92\% & 25\% & 9\% & $\sim$10$\times$ \\
\textbf{Compute SOL, device-wide}     & \textbf{92.0\% (mean), 97\% (peak)} & \textbf{22.6\%} & \textbf{7.9\%} & $\sim$\textbf{11.7}$\times$ \\
Wave efficiency (mean)                & 99.8\% & 85.5\% & 80.5\% & $\sim$1$\times$ \\
\gls{gmma} $M$-fragment fill $\eta_M$ (dense GEMM mean) & 100\% & 70.2\% & 5.86\% (1.6--12.5\%) & $\sim$17$\times$ \\
L2 fit ratio $\rho$ (mean dense; $\rho{>}1$ spills) & 14.7$\times$ & 16.0$\times$ & 2.0$\times$ & $\sim$8$\times$ smaller (still spills 2$\times$) \\
Analytical launch overhead (Llama-3-8B) & 1.3\% & 28.5\% & 29.9\% & $\sim$23$\times$ \\
\texttt{long\_scoreboard} \% of stalls & 19--36\% & 72--83\% & 70--84\% & $\sim$3$\times$ \\
\bottomrule
\end{tabular}
\end{table*}

\subsection{Per-token time decomposition}\label{sec:pertoken}

Kernel-level bottlenecks matter in proportion to their per-token
time share.  In the main \(B{\le}8\) sweep, the FFN is the dominant
cost at 58--66\% mean across models and regimes (the \(B{=}32\)
validation arm shifts toward \gls{fa3}; \S\ref{sec:b32}).
For example, Llama-3-8B decode at $(S{=}2048,B{=}1)$ takes
5.27\,ms per token, of which 63.7\% is FFN (\texttt{gate\_up}
41\%, \texttt{down\_proj} 22\%), 30.8\% attention, 3.2\%
norm/embed, and 2.2\% reduce.  FFN is therefore the natural
denominator for ``X\% improvement'' claims, including the
compute-SOL, wave, and launch-overhead fixes of \S\ref{sec:implications}.

%==============================================================================
\section{Discussion}\label{sec:discussion}

\smartparagraph{Implications.}\label{sec:implications}
The findings translate into levers at three layers.
\emph{Software and serving}: batch to the compute--memory crossover
rather than to ``more batch'' --- $B^{*}{\in}[25,49]$
(\S\ref{sec:ridge}) bounds where dense decode leaves the
bandwidth-bound regime, a concrete target for batching, speculative
decoding~\cite{specdec}, and PD-disaggregation; and schedule decode and
warm prefill separately, since warm prefill is closer to decode than to
cold prefill (\S\ref{sec:tc}).  Decode's analytical launch overhead
reaches $\sim$30\% of the iteration (\Cref{tab:cross-metric}), so a
dispatch-amortizing decode path (graph capture or a persistent kernel) helps.

\noindent\emph{Libraries and code generation}: the largest decode losses are
not retiling problems --- no \texttt{cuBLASLt} tile yields a
132-multiple grid for decode-QKV (\S\ref{sec:wave}), warp-pool
headroom needs register \emph{and} SMEM cuts \emph{together}
(\S\ref{sec:rs-bind}), and the \gls{gmma} $m64$ floor caps every
decode cell at $\eta_M{=}B/64$ regardless of tile choice
(\S\ref{sec:tc}).  It is worth separating two levers that are easily
conflated.  A persistent decode kernel removes launch overhead and
wave-quantization loss, but it leaves $\eta_M$ unchanged, because it adds no
token rows.  Continuous batching does not raise $\eta_M$ either, since it
already sets the decode-GEMM row count to $M{=}B$ (\S\ref{sec:regimes}) and
the latency target holds $B$ below the crossover of \S\ref{sec:ridge}.  Rows
can only come from speculative, tree, or multi-token decoding, which give
$M{=}kB$ and fill the fragment at $k{=}8$, $B{=}8$.

\noindent\emph{Architecture}: decode stresses memory and host
dispatch, not the tensor pipe, raising the open question of whether
finer-grained power/clock gating of the tensor datapath is warranted;
and decoupling the register file from shared memory would let one
resource give occupancy headroom instead of both having to move.  Finally, operators should report
multiple denominators: a kernel can look healthy on SM-time and warp
axes while delivering a small fraction of peak matmul throughput.

\smartparagraph{Threats to validity.}\label{sec:threats}
Six caveats qualify our claims.
\cnum{1}~\emph{\glsentryshort{fa3} row blending}: the standard \flashattn{}
row blends \gls{fa3} main and combine, so reading it as main alone
underestimates decode \gls{fa3} duty (we re-read the two separately
from the \texttt{.ncu-rep} files; \S\ref{sec:method}).
\cnum{2}~\emph{Linear $B^{*}$ extrapolation} is validated and refined
by the $B{=}32$ arm (\S\ref{sec:b32}: Qwen3-14B crosses near
$B^{*}{\approx}31$, Llama-3-8B approaches, Qwen3-32B crosses at short
$S$ but is KV-capped at long $S$, Qwen3-30B-A3B shifts to a
routed-FFN/\glsentryshort{fa3} co-bottleneck) --- a diagnostic, not a predictor.
\cnum{3}~\emph{Single-GPU, TP=1, single H100 SKU}: TP alters per-GEMM
shape and adds all-reduce, and HBM-bound numbers may shift on other
H100 SKUs.
\iffalse % Threats item depended on commented-out \S5.12
\cnum{4}~\emph{The MoE proxy is warm-prefill-only}: the within-slot
signal (\S\ref{sec:moe}) is consistent with routing imbalance but,
given early-layer outliers (gate timings $\sim$1.5$\times$ higher
in layers 0--5), is not strictly separable from L2/weight-warmup
transients without a balanced-router replay; decode and cold prefill
sit at the noise floor.
\fi
\cnum{4}~\emph{Launch-overhead constant}: the analytical $\tau{=}5\,\mu$s
is host-/driver-dependent and captures raw dispatch only, not
runtime-path/replay/scheduling cost.
\cnum{5}~\emph{Working-set model is first-order}: no scratch/tile
staging; within $\sim$10--20\% of \texttt{dram\_\_bytes} on
cross-checked kernels.
\cnum{6}~\emph{Warm-prefill cache state}: synthetic warmup yields
$\geq$99\% prefix-cache hit; realistic multi-turn intermediate hits
are uncharacterized.

%==============================================================================
\section{Related Work}\label{sec:related}

% \smartparagraph{Serving stacks.}  We characterize the workload that
% vLLM/PagedAttention~\cite{vllm}, TensorRT-LLM~\cite{tensorrtllm}, SGLang~\cite{sglang},
% FlashAttention 1/2/3~\cite{flashattn,flashattn2,flashattn3}, Orca~\cite{orca},
% Sarathi-Serve~\cite{sarathi}, DistServe~\cite{distserve}, and Splitwise~\cite{splitwise}
% run, rather than proposing a new serving system.

\smartparagraph{GPU characterization and roofline.} Prior LLM inference work on A100/H100~\cite{llminfer-a100} measures end-to-end performance. We instead map per-role kernels to NCU counters, validate with calibration, and tie analysis to measurement via a controlled \texttt{cuBLASLt} tile-space enumeration. Hopper microbenchmarking~\cite{hopper-training} studies primitives in isolation; we show how an end-to-end serving stack (vLLM/\gls{fa3}/\texttt{cuBLASLt}) composes them. Roofline~\cite{roofline,hroofline,llm-roofline} is the main analytical precedent; we extend it with wave quantization, kernel-selection stability, and a \texttt{cuBLASLt} algorithm-space enumeration. Our counter-level results (the \glsentryshort{fa3} blending caveat and the register/shared-memory occupancy co-limit) refine the meaning of NCU metrics~\cite{ncu-doc}.
FlashDecoding++~\cite{flashdecodingpp} also treats the small-$M$ decode GEMM as
a first-order problem and builds a flat-GEMM kernel that pads $M$ to 8 instead
of 64.  The two efforts meet from opposite directions.  That work proposes a
kernel; we measure what the stock library path actually does and show why the
row deficit survives every tile the library can produce
(\S\ref{sec:wave}).  Read together, our enumeration explains why a custom
kernel is the level at which the problem has to be solved.

\smartparagraph{MoE, profiling, and benchmarks.} MoE work is mostly training-focused, covering routing, expert parallelism, load balancing, grouped-GEMM kernels, and distributed-MoE imbalance~\cite{gshard,switchtrans,expertpar,routebalance,megablocks,moetrain}. Benchmark suites such as MLPerf~\cite{mlperf} report end-to-end latency or throughput; we instead report per-role microarchitectural metrics.

%==============================================================================
\section{Conclusion}
  \label{sec:conclusion}
  SM utilization is a useful number, but it is not a diagnosis. In LLM
  inference, the same kernel can look busy or idle depending on whether we
  count elapsed SM activity, tensor-pipe work, launch-grid coverage,
  occupancy headroom, or useful GMMA fragment fill, and collapsing these
  views into one scalar hides where performance is actually lost. Decode
  makes this concrete. Dense decode GEMMs often reach most of the chip,
  yet at small batch sizes Hopper's \gls{bf16} \gls{gmma} \texttt{m64}
  fragment floor leaves most rows of every instruction empty, so SM-busy
  stays high while useful matmul does not. This is an instruction-set
  ceiling, not a tiling choice, and no \texttt{cuBLASLt} retuning can lift
  it. \gls{fa3} decode loses time to a different cause, barrier-style
  waits rather than memory stalls. Different bottlenecks need different
  fixes: small-batch decode is improved less by retuning \texttt{cuBLASLt}
  tiles than by raising the effective row dimension $M$, through
  cross-request token packing or persistent decode kernels. More broadly,
  LLM inference studies should report counter-pinned, per-role metrics
  rather than a single device-wide utilization percentage.
%==============================================================================
\section*{Acknowledgments}
We would like to thank the anonymous reviewers for their insightful
comments and suggestions on this paper.  This work has been partially
supported by Vinnova (Sweden's Innovation Agency), the Swedish Research
Council (agreement No.\ 2021-04212), KTH Digital Futures, and the Knut
and Alice Wallenberg Foundation (Wallenberg Scholar Grant for
Prof.\ Dejan Kosti\'c).

%==============================================================================
% \bstctlcite activates the @IEEEtranBSTCTL entry in reference.bib,
% which disables the repeated-author dash so each entry prints its
% author in full.
\bstctlcite{BSTcontrol}
\bibliographystyle{IEEEtranS}
\bibliography{reference}

\end{document}